\documentclass[aps,prl,reprint,twoside,floatfix,superscriptaddress,showpacs,showkeys]{revtex4-1}

\usepackage{graphicx}
\usepackage{amsmath}
\usepackage{amssymb}
\usepackage{amsfonts}
\usepackage{dcolumn}
\usepackage{epsfig}
\usepackage{subfigure}
\usepackage{bm}
\usepackage[version=3]{mhchem}
\usepackage{textcomp}
\usepackage{miller}
\usepackage{gensymb}

\newcommand{\Wt}{\ensuremath{W_\mathrm{t}}}
\newcommand{\Wopt}{\ensuremath{W_\mathrm{opt}}}

\begin{document}

\title{Origin of traps and charge transport mechanism in hafnia}

\author{D.~R.~Islamov}\email{damir@isp.nsc.ru}
	\affiliation{Rzhanov Institute of Semiconductor Physics,
		Siberian Branch of Russian Academy of Sciences,
		Novosibirsk, 630090, Russian Federation}%
	\affiliation{Novosibirsk State University,
		Novosibirsk, 630090, Russian Federation}%
\author{V.~A.~Gritsenko}\email{grits@isp.nsc.ru}
	\affiliation{Rzhanov Institute of Semiconductor Physics,
		Siberian Branch of Russian Academy of Sciences,
		Novosibirsk, 630090, Russian Federation}%
	\affiliation{Novosibirsk State University,
		Novosibirsk, 630090, Russian Federation}%
\author{C.~H.~Cheng}
	\affiliation{Dept. of Mechatronic Technology, National Taiwan Normal University, Taipei, 106, Taiwan ROC}%
\author{A.~Chin}\email{albert\_achin@hotmail.com}
	\affiliation{National Chiao Tung University, Hsinchu, 300, Taiwan ROC}%

\date{\today}

\begin{abstract}
In this study, we demonstrated experimentally and theoretically
that oxygen vacancies are responsible for the 
charge transport in \ce{HfO2}.
Basing on the model of phonon-assisted tunneling between traps,
and assuming that the electron traps are oxygen vacancies,
good quantitative agreement between the experimental
and theoretical data of current-voltage characteristics were achieved.
The thermal trap energy of $1.25$\,eV in
\ce{HfO2} was determined based on the charge transport experiments.
\end{abstract}



\maketitle
Knowledge about charge transport mechanisms
hafnia (hafnium oxide, \ce{HfO2}) is
crucial for modern microelectronics,
because 
high-$\kappa$ \ce{HfO2} is used as a gate dielectric
in high-speed MOSFETs  \cite{Ma:JDMR5:36, Robertson:RPP69:327}
and FinFETs \cite{HfO2:FinFET:SSE51:285, HfO2:FinFET:SST24:125001}
and a blocking insulator in
\ce{Si}-oxide-nitride-oxide-silicon-type (SONOS) flash memory cells
\cite{HfO2:SONOS:NanoTech10:260, HfO2:SONOS:JEDL30:775}.
Hafnia is a promising candidate
to used sa active medium in resistive random access memory,
which would involve combining the most favorable properties of
both high-speed dynamic random access memory and non-volatile flash memory \cite{rram:jjyang, Goux:APL97:243509, PhysRevB.85.195322}.
However, an unresolved physics is the nature of defects and traps
that are responsible for the charge transport in \ce{HfO2}.
The atomic structure of defect that affects the localization
and charge transport still remains unclear.
Currently, the accepted hypothesis is that oxygen vacancies
are responsible for charge transport in dielectric.
Although many studies have investigated the theory of the atomic
and electronic structure of oxygen vacancies in hafnia
\cite{Guha:PRL98:196101, Ramo:PRL99:155504, Foster:PRB65:174117, PhysRevB.75.104112, Ramo:PRB75:205336, PhysRevB.81.085119, Xiong:APL87:183505, Broqvist:APL89:262904},
direct experimental data regarding the presence of oxygen vacancies in hafnia
were reported recently \cite{Perevalov:APL104:071904}.
It was shown that oxygen vacancies
in hafnia are responsible for blue luminescence band at
$2.7$\,eV and a luminescence excitation band at $5.2$\,eV,
and a hypothesis that the oxygen vacancies in hafnia act as traps
in charge transport through the dielectric
was discussed \cite{Perevalov:APL104:071904}.
In this case thermal energy traps in \ce{HfO2} is equal to
a half of the Stokes luminescence shift
$\Wt = (5.2-2.7) / 2 = 1.25$\,eV. 

A lot of charge transport studies described experiment
results by Pool-Frenkel (PF) mechanism
in hafnia-based structures \cite{HfO2:PF:JAP110:084104, HfO2:PF_1.2eV:JAP102:073706, Zhu:JEDL23:97}.
However, the most investigations
explained their results qualitatively, and did not
get quantitative agreement of the phenomenological parameters
such as dynamic permittivity, trap energy, frequency factor etc.
The most part of transport investigations did not get into account
neither charge trap density,
 which depends on thin film fabrication technology,
nor phonon influence on electron and hole transport,
 which might be significant at high temperatures.
 
In this letter, we phonon-assisted tunneling between traps
conduction mechanism in \ce{HfO2} was developed
with good quantitative as well as qualitative agreement. 
It was clearly shown that oxygen vacancies are responsible
for the charge transport
in \ce{HfO2} and \ce{HfO2}-based devices. 

Transport measurements were performed for
metal-in\-su\-la\-tor-semiconductor (MIS) and
metal-insulator-metal (MIM) structures.
For the MIS \ce{Si}/\ce{HfO_x}/\ce{Ni} structures,
the 20-nm-thick amorphous hafnia was deposited on a $n$-type Si \hkl(100) wafer
by using the atomic layer deposition (ALD) system.
Tetrakis dimethyl amino hafnium (TDMAHf) and water vapor
were used as precursors at a chamber temperature of $250\degree$C
for \ce{HfO_x} film deposition.

Another set of MIS samples with 8-nm-thick \ce{HfO_x} films
was fabricated
using physical vapor deposition (PVD).
A pure \ce{HfO2} target was spattered by an electron beam,
and \ce{HfO2} were deposited on the
$n$-type Si \hkl(100) wafer.
Low temperature post-deposition annealing (PDA)
during $15$\,min at $400\degree$C was applied
to prevent the growth
of interfacial \ce{SiO_x} \cite{Lin:IEEE_J_EDL30:999}.
Structural analysis shows that the resulting
\ce{HfO_x} films were amorphous.
To fabricate \ce{Si}/\ce{TaN}/\ce{HfO_x}/\ce{Ni}
MIM structures, we deposited the 8-nm-thick amorphous hafnia
on 100-nm-thick TaN films on \ce{Si} wafers,
using PVD.
We did not apply any post-deposition annealing
to produce the most non-stoichiometric films.
All samples for transport measurements were equipped with round
50-nm-thick Ni gates with a radius of $70$\,$\mu$m.
Transport measurements were performed using a Hewlett Packard 4155B
semiconductor parameter analyzer and an Agilent E4980A precision LCR meter. 

\begin{figure}
  \includegraphics[width=\columnwidth]{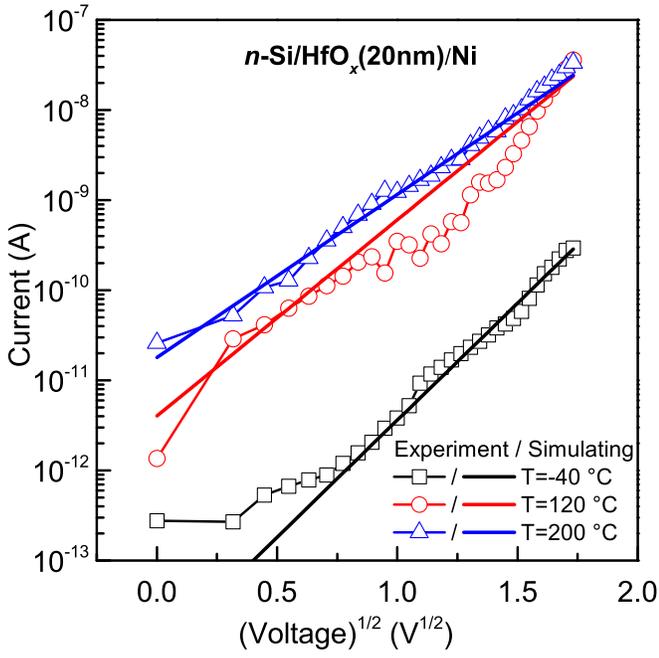}
  \caption{(Color online)
      Experimental current-voltage characteristics (characters)
      of $n$-\ce{Si}/\ce{HfO_x}/\ce{Ni} MIS(PVD) structure
      and simulation (lines)
      by Frenkel model of the trap ionization
      (\ref{eq:current}), (\ref{eq:frenkel})
      at different temperatures on PF plot.
  }
  \label{f:PVD_MIS_IVT_Frenkel}
\end{figure}

The experimental current-voltage ($I$-$V$) characteristics in
MIS(PVD) structures,
measured at different temperatures $T$ with a positive bias applied to the
\ce{Ni} contact, are shown in Fig.~\ref{f:PVD_MIS_IVT_Frenkel},
graphed by different characters in PF ($\log(I)$-$\sqrt{V}$) plot.
The current grows exponentially with increasing of the gate
voltage and temperature.
We attempted to explain the experimental data
by using isolated trap ionization model:
\begin{equation}
  J=eN^{2/3}P,
  \label{eq:current}
\end{equation}
where $J$ is the current density,
$e$ is the electron charge,
$N$ is the charge trap density,
$P$ is the probability of trap ionization
per second, and has different dependencies on
electric field $F$ and temperature.
In term of PF model the probability is
\begin{equation}
  P=\nu\exp\left(-\frac{\Wt-\beta_\mathrm{PF}\sqrt{F}}{kT} \right).
  \label{eq:frenkel}
\end{equation}
$\nu$ is the frequency factor which was defined as
$\nu\simeq\Wt/h$,
$\Wt$ is thermal trap energy 
(the energy of thermal ionization of the trap),
$h$ is the Planck constant,
$\beta_\mathrm{PF}$ is Poole–Frenkel coefficient,
is the electric field,
and $k$ is the Boltzmann constant \cite{PhysRev:54:647}.
Experimental $I$-$V$ characteristics and results of
the fitting procedure are shown in Fig.~\ref{f:PVD_MIS_IVT_Frenkel}.
As can be seen, Frenkel model (\ref{eq:frenkel})
describes the experiment data
qualitatively very good. However, quantitative fitting procedure
returns nonphysical fitting parameter values:
the slopes of the fitting lines with Poole–Frenkel coefficient
$
  \beta_\mathrm{PF} = {e^3}/{\pi\varepsilon_0\varepsilon_\infty}
$
give the dynamic permittivity $\varepsilon_\infty(T)=10 \div 20$,
which is much higher than $\varepsilon_\infty(\ce{HfO2})=4$.
$\varepsilon_0$ is vacuum permittivity (dielectric constant).
Further fittings return
$N\sim4$\,cm$^{-3}$ (!) and $\Wt=0.3\div 0.4$\,eV.
Found values the charge trap density of $N\sim4$\,cm$^{-3}$
at $\nu\simeq\Wt/h\sim 10^{14}$\,s$^{-1}$
corresponds to one trap per $\sim 2600$ \ce{Ni} contacts,
thus this is unrealistic value.
This Taking these into account
it was concluded that despite the fact that PF model
describes the experiment data qualitatively,
there is no quantitative agreement between experiments and theory.
We tried to describe our experiments with other charge 
transport models in dielectrics;
Hill model of overlapped traps ionization \cite{PhilMag:23:59},
and the model of multiphonon trap ionization \cite{PhysRevB:25:6406}.
However, the fitting procedures involved in these models
returned the nonphysical fitting parameter values
as well as PF model.

\begin{figure}
  \includegraphics[width=\columnwidth]{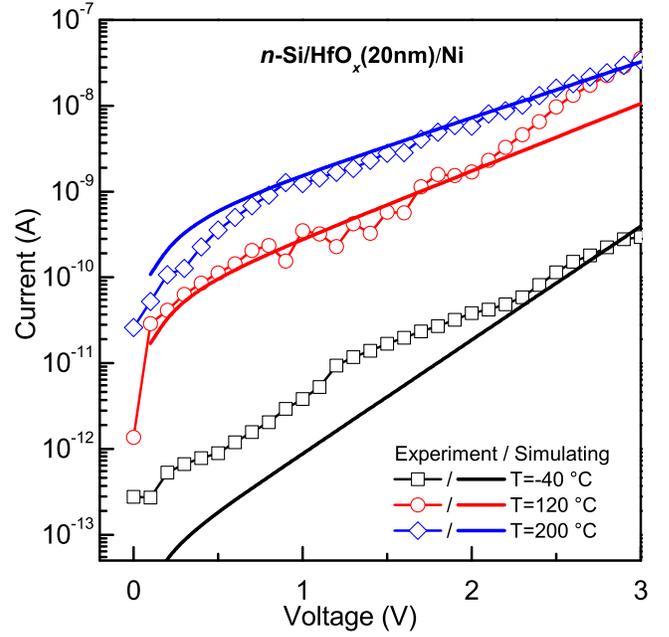}
  \caption{(Color online)
    Experimental current-voltage characteristics (characters)
    of $n$-\ce{Si}/\ce{HfO_x}/\ce{Ni} MIS(PVD) structure
    and simulation (lines)
    by phonon-assisted tunneling between traps
    (\ref{eq:nasyrov})
    at different temperatures.
  }
  \label{f:PVD_MIS_IVT_Nasyrov}
\end{figure}

To describe the experiments quantitatively and qualitatively,
we performed simulations based on the model of phonon-assisted
tunneling between traps \cite{JAP:109:093705}. In this model
the probability of electron tunneling between traps per second
is defined as following: 
\begin{equation}
\begin{aligned}
  P=&\frac{\sqrt{2\pi}\hbar \Wt}{m^* s^2 \sqrt{\Wopt-\Wt}}
  \exp\left(-\frac{\Wopt-\Wt}{2kT} \right)\times\\
  \times & \exp\left(-\frac{2s\sqrt{m^* \Wt}}{\hbar} \right)
  \sinh\left(\frac{eFs}{2kT} \right),
  \label{eq:nasyrov}
  \end{aligned}
\end{equation}
$\hbar=h/2\pi$,
$\Wopt$ is the energy of optical excitation of the trap,
$m^*$ is the effective mass,
$s = N^{-1/3}$ is mean distance between traps.
The results of this multi-parameter fitting procedure are shown in
Fig.\ref{f:PVD_MIS_IVT_Nasyrov}, graphed in solid lines.
This procedure yielded the values of different transport parameters,
$N=6.8\times10^{19}$\,cm$^{-3}$,
$\Wt=1.25$\,eV,
$\Wopt=2\Wt=2.5$\,eV,
and $m^*=(0.3\div 0.4)m_\mathrm{e}$
($m_\mathrm{e}$ is a free electron mass).
Quantitatively, there is full agreement between the model
of phonon-assisted tunneling between traps and the experimental data.
The trap thermal energy value of $1.25$\,eV that was obtained is close
to that of $1.2$\,eV \cite{Takeuchi:JVSTA22:1337} and 
$\Wt=1.36$\,eV  \cite{Jeong:PRB71:165327} observed earlier,
and equal to a half of the Stokes luminescence shift \cite{Perevalov:APL104:071904}.
Furthermore, the trap optical energy value of $\Wopt=2.5$\,eV
is close to the calculated value of $2.35$\,eV
for the negatively charged oxygen vacancy in hafnia reported
earlier \cite{Ramo:PRB75:205336}.

\begin{figure}
  \includegraphics[width=0.5\columnwidth]{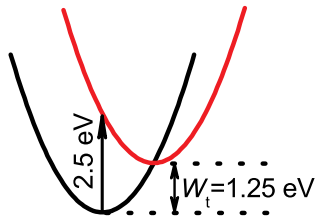}
  \caption{(Color online)
   Configuration coordination energy diagram of trap ionization process
   on negative charged oxygen vacancy in hafnia
   (lower term is filled ground state,
   upper term is excited empty state).}
  \label{f:configdiag:thermal}
\end{figure}

Fig.~\ref{f:configdiag:thermal} shows the configuration diagram of
a negatively charged oxygen vacancy (electron trap) in hafnia.
A vertical transition with a value of $2.5$\,eV corresponds
to the optical trap excitation,
transitions of $1.25$\,eV correspond to thermal trap energy.

\begin{figure}
  \includegraphics[width=\columnwidth]{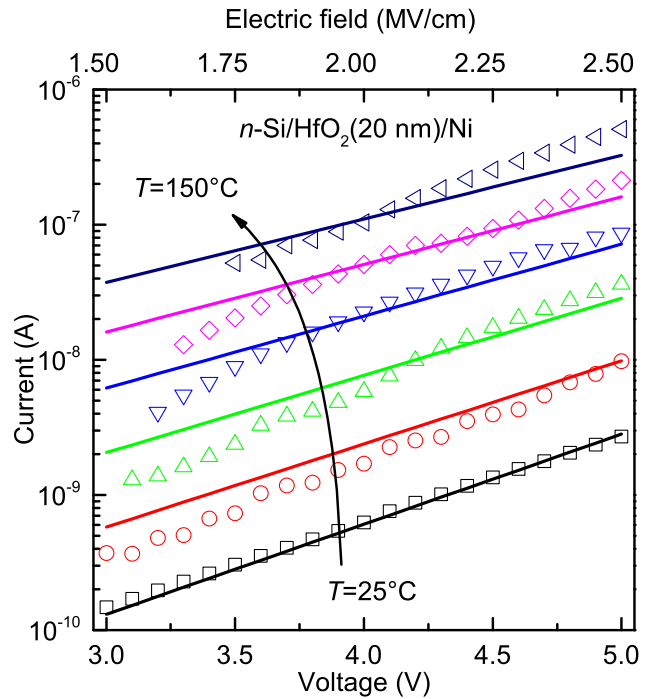}
  \caption{(Color online)
    Experimental current-voltage characteristics (characters)
    of $n$-\ce{Si}/\ce{HfO_x}/\ce{Ni} MIS(ALD) structure and simulation (lines)
    by phonon-assisted tunneling between traps
    (\ref{eq:nasyrov})
    at different temperatures.
  }
  \label{f:ALD_MIS-IVT}
\end{figure}

The same procedure was applied to experiment data
of the charge transport measurements in MIS(ALD) and MIM structures.
Experiment current-voltage characteristics compared with
simulations in terms of the model
of phonon-assisted tunneling between traps in MIS(ALD)
are shown in Fig.~\ref{f:ALD_MIS-IVT}.
Fitting procedure returns the following parameters values:
$N=2.5\times10^{20}$\,cm$^{-3}$,
$\Wt=1.25$\,eV,
$\Wopt=2\Wt=2.5$\,eV,
and $m^*=0.8m_\mathrm{e}$.
Different values of fitting parameters of
MIS(ALD) and MIS(ALD) structures have only
the trap density $N$ and effective mass $m^*$.
The difference of effective mass values is
due to bulk space charge
(due to captured in traps electrons and holes),
which is adequately addressed in \cite{Nasyrov:JAP105:123709}.
Neither thermal trap energy nor optical trap energy depend on
film fabrication technology.

\begin{figure}
  \includegraphics[width=\columnwidth]{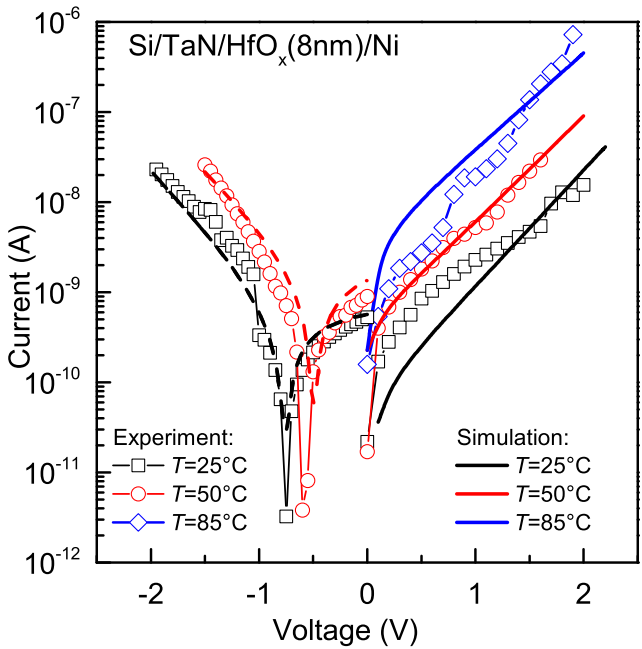}
  \caption{(Color online)
    Experimental current-voltage
    characteristics (characters) in
    \ce{Si}/\ce{TaN}/\ce{HfO_x}/\ce{Ni} MIM structures
    at different temperatures.
    Black, red and blue solid lines represent $I$-$V$
    simulations by phonon-assisted tunneling between traps
    at positive bias. Dashed lines show
    $I$-$V$ simulations taking into account
    displacement current (\ref{eq:I_D}).
  }
  \label{f:MIM_IVT}
\end{figure}

Fig.~\ref{f:MIM_IVT} shows the experimental data
of current-voltage measurements in MIM structures
at different temperatures.
The solid lines present the results of simulations
in terms of phonon-assisted tunneling between traps
(\ref{eq:nasyrov}).
MIM structures have the following parameter values:
$N=5.5\times10^{20}$\,cm$^{-3}$,
$\Wt=1.25$\,eV,
$\Wopt=2\Wt=2.5$\,eV,
and $m^*=0.9m_\mathrm{e}$.

An artifact feature
of experimental $I$-$V$-$T$ curves that the zero current
is observed at non zero but negative voltages
as shown in Fig.~\ref{f:MIM_IVT}. This phenomenon is due to
displacement current
\begin{equation}
  I_{D}= C\cdot {dV}/{dt},
  \label{eq:I_D}
\end{equation}
$C$ is capacity of the sample,
$dV/dt=+0.3$\,V$/$s is voltage sweep rate.
Taking into account (\ref{eq:I_D}) in simulation of
$I$-$V$-$T$ characteristics
(\ref{eq:current}), (\ref{eq:nasyrov})
with found fitting parameters
the artifact feature is described with
good agreement as shown by dashed lines in Fig.~\ref{f:MIM_IVT}.

The difference between different MIS ad MIM
structures in effective mass is
due to bulk space charge.
However, it is important to notice that the trap's energy parameters
are invariants of grown structures and film fabrication techniques.
Consequently, we found that the nature of charge carrier transport
in hafnia and hafnia-based structures
is phonon-assisted tunneling between traps.
This charge transport model is more simple than based in
quasi-continuous spectra of charge trap energy $1.4-2.4$\,eV,
proposed by L. Vandelli \textit{et al.} \cite{Vandelli:ED58:2878}.

These results combined with spectra measurements and quantum chemical
simulations \cite{Perevalov:APL104:071904}
show that namely oxygen vacancies are responsible for charge
transport in \ce{HfO_x}, and the oxygen vacancies play the role
or charge traps.

Previous experiments in charge transfer have demonstrated that hafnia conductivity
is bipolar (or two-band) \cite{APL201108, Ando:EDL32:865, Vandelli:ED58:2878}:
electrons are injected from a negatively shifted contact in the dielectric,
and holes are injected from a positively shifted electrode in the dielectric.
For the reason of simplicity, the current study took into account
electron conductivity only.

To summarize, we examined the transport mechanisms of \ce{HfO2},
demonstrating that transport in hafnium oxide is described by the model
of phonon-assisted tunneling between traps.
Simulating the current-voltage characteristics of this model
and comparing experimental data with calculations revealed the energy
parameters of the traps in hafnia:
the thermal trap energy of $1.25$\,eV
and the optical trap energy of $2.5$\,eV.
Phonon-assisted tunneling between traps charge transport model
describe experiment data results
with excellent qualitative and quantitative agreement,
while standard PF model has qualitative agreement only with
unrealistic values for material parameters.
These results jointly with earlier ones \cite{Perevalov:APL104:071904}
facilitated determining that oxygen vacancies act
as charge carrier traps.

Our results can be used to predict the leakage currents in
\ce{HfO2}-based devices and applications.
High-quality MOSFET and FinFET
transistors and
SONOS flash memory 
require
low leakage currents through the gate dielectrics and
blocking insulator, while
different states in resistive memory cells
must be distinguishable over a wide range of temperatures.
Temperature dependence of memory window
(resistance ratio in different states)
in resistive memory
might be predicted as well.

This work was particularly supported by National Science Council, Taiwan
(grant No.~NSC103-2923-E-009-002-MY3)
(growing test structures, preparing samples, performing transport measurements),
and by the Russian Science Foundation (grant No.~14-19-00192)
(calculations, modeling).

\bibliographystyle{apsrev}
\bibliography{IEEEabrv,../../../bibtex/Technique,../../../bibtex/GeO2,../../../bibtex/TiO2,../../../bibtex/HfO2,../../../bibtex/SiO2,../../../bibtex/TaOx,../../../bibtex/Theory,../../../bibtex/percollation,../../../bibtex/Memristor,../../../bibtex/computing}

\end{document}